# Share Price Prediction of Aerospace Relevant Companies with Recurrent Neural Networks based on PCA

Linyu Zheng[a,] and Hongmei He[b,1]

[a]AVIC General Huanan Aircraft Industry Co., Ltd., Aviation Industry Zone, Sanzao Town, Jinwan District, Zhuhai City, Guangdong Province, China, 519040

[b]School of Aerospace, Transport and Manufacturing, Cranfield University, Cranfield, UK, MK43 0AL[2]

**ABSTRACT**

The capital market plays a vital role in marketing operations for the rapid development of the aerospace industry. However, due to the uncertainty and complexity of the stock market and many cyclical factors, the stock prices of listed aerospace companies fluctuate significantly. This makes the share price prediction challengeable. To improve the prediction of share price for aerospace industry sector and well understand the impact of various indicators on stock prices, we provided a hybrid prediction model by the combination of Principal Component Analysis (PCA) and Recurrent Neural Networks.

We investigated two types of aerospace industries (manufacturer and operator). The experimental results show that PCA could improve both accuracy and efficiency of prediction. Various factors could influence the performance of prediction models, such as finance data, extracted features, optimisation algorithms, and parameters of the prediction model. The selection of features may depend on the stability of historical data: technical features could be the first option when the share price is stable, whereas fundamental features could be better when the share price has high fluctuation. The delays of RNN also depend on the stability of historical data for different types of companies. It would be more accurate through using short-term historical data for aerospace manufacturers, whereas using long-term historical data for aerospace operating airlines.

The developed model could be an intelligent agent in an automatic stock prediction system, with which, the financial industry could make a prompt decision for their economic strategies and business activities in terms of predicted future share price, thus improving the return on investment. The study is for the prediction of aerospace industries at pre-COVID-19 time. Currently, COVID-19 severely influences aerospace industries. The developed approach can be used to predict the share price of aerospace industries at post COVID-19 time.

---

[1] Corresponding author: Dr Hongmei He is with Cranfield University, h.he@cranfield.ac.uk

# 1. Introduction

As part of the transportation industry, the aerospace industry has become an indispensable part of human life. Aerospace is the most active and influential field of science and technology in the 21st century. The outstanding achievements in this field mark the level of development of human civilisation. It also reflects the level of comprehensive national strength and science and technology of a country.

In the aerospace sector, the most representative industries are aerospace manufacturers (e.g. Boeing, Airbus) and aerospace operating sector (e.g. Alaska Airlines, United Airlines). To expand financing, invest in new projects or new businesses, and increase market competitiveness, many companies choose to be listed on stock market. This is because that entering the stock market is a straightforward approach to raising more capital for aerospace companies, expanding international market share, gaining higher public trust, and creating more transparent and measurable company values.

The price of stock markets is the critical indicator of economic growth in a country. In 2015, there were more than 60 stock exchanges in the world. At the mid-2017, the total stock market value size in the world was around $76.3 trillion US dollars. The stock price prediction is trying to confirm the future values of the stocks or other financial tools traded in the stock exchanges (Malkiel, 1991). The investors and traders are interested in share price prediction because accurate prediction may produce high profit through stock trading. However, the share prices of aerospace companies fluctuate significantly due to many cyclical and non-cyclical factors. Each factor has a significant impact on profit, but these elements are not synchronised. Due to the uncertainty and complexity of the stock market, the prediction of share prices is a critical challenge, and thus the stock trading is a high-risk financial activity. Therefore, a slight improvement of share price prediction could bring a substantial return of rate for investors.

Although financial market analysis requires knowledge, intuition, and experience, the automation process has been growing steadily because of the availability of large finance data and the development of machine learning technology. There are two conventional methods for stock price prediction: fundamental analysis and technical analysis. Fundamental analysis is to concern the stock itself (Harrington, 2003), while technical analysis is to predict the future trend of the transaction values and volumes by researching the past market data without concerning any fundamental aspect and financial condition ratios of the company itself (Kirkpatrick & Dahlquist, 2015).

Correspondingly, the two sets of data, used in the two analysis methods, are called fundamental stock data and technical stock data, respectively. Moreover, various machine-learning techniques have been widely used in the finance industry.

This research is to improve the accuracy of share price prediction in aerospace industry using advanced machine learning techniques, to provide decision-makers with references about their stock price as well as various factors that could affect the stock price in future for their economic strategies and business activities. We develop a hybrid approach with two stages of PCA and recurrent neural network for stock price prediction. Two types of aerospace industries are studied based on their history data. One is a manufacturing company, denoted as MFG, and the other is an operating company, denoted as OPR. We select 15 fundamental features and 15 technical features, respectively. The three groups of experiments will be conducted based on the different data types: fundamental data, technical data and their combination.

The rest of the paper is organised as follows: Section 2 provides a brief literature review on the research of computational finance, especially the methods and techniques on share price analysis and prediction. Section 3 describes the hybrid model for share price prediction. Section 4 provides the experiments and comprehensive evaluation. Finally, Section 5 provides conclusions and future work.

**2. Existing Work**

*2.1. Fundamental and Technical Analysis for Stock Price Prediction*

Fundamental analysis is the process of looking at a business at the most basic or fundamental financial level. It examines the key ratios of a business to determine its financial health. Fundamental analysis can provide an idea of the value of what a company's stock should be. Hence, fundamental analysis concerns the stock itself, such as the assets, liabilities, and incomes of the target companies by analysing their financial statements, as well as the ratios of previous performance, such as Price-to-Earnings Ratio (P/E ratio) (Harrington, 2003), which is the ratio for valuing a company that measures its current share price, related to its earnings per-share (EPS). Fundamental analysis is based on the belief in the business needs for the capital to keep operating. If the company runs well, it should obtain additional capital awards which will lead to stock price soared (Fundamental Focus, 2011). Fundamental analysis is conducted from the global economy firstly and then national economy before analysing a specific industry and a specific company. It is a top-down process. As the fundamental analysis is a relatively reasonable and objective method, it is used extensively (Harrington, 2003).

Technical analysis is to predict the future stock price and makes the trade decisions of financial derivatives based on the position and the seasonal change theory of the market trend. The principles of technical analysis are that the price will reflect all the relevant information of the market and the history trend will repeat itself (Edwards et al., 2012).

Both analyses can predict stock price to some extent. The main difference between these two methodologies is that the datasets of fundamental analysis could update very slow. For example, return on assets (ROA) and return on equity (ROE) are likely to upgrade every three months, rather than daily as with technical analysis dataset, e.g. stock price data (Larkin, 2014).

For the fundamental analysis, there are lots of indicators in stock analysis processes. 42 features based on the fundamental analysis were extracted by different analysts for the prediction of share price. The 5 feature vectors, such as profitability, growth, liquidity, solvency and operational efficiency, were used for tackling the value stock analysis of IT stocks in Taiwan. The profitability vector includes ROA, operational gross profit, operational profit, net profit after tax; the growth vector includes net profit after tax growth rate, ROA growth rate, total assets growth rate, revenue growth rate, gross profit growth rate; the liquidity vector includes quick ratio, liquidity ratio, cash ratio; the solvency vector includes debt ratio, interest coverage ratio; and operational efficiency vector includes asset turnover rate, inventory turnover rate, average days for sales (Shen & Tzeng, 2015a). Besides, the rate of return, the number of transactions, gross profit, gross loss, the number of profitable transactions, the number of consecutively profitable transactions, the number of unprofitable consecutive transactions, Sharpe ratio, the average coefficient of volatility, the average rate of return per transaction, the value of the evaluation have been used in the decision system on FOREX market (Korczak et al., 2016). In order to select optimal stocks from the stock market and predict the future price trends, Chen et al. (2017) developed an improved fundamental approach with 14 features that were widely used as indicators in the analysis of finance and investment, such as long-term funds to fixed assets, current ratio, interest guarantee, average inventory turnover, average collection turnover, fixed assets turnover, total assets turnover, return on total assets, total stockholders' equality, operating income to capital, pre-tax income, net income, earnings per share, and the price-earnings ratio. More features can be extracted from the daily stock market data. Moving averages and relative strength index (RSI) have been used in a mobile app to classify stock market (Larkin, 2014). To support investment decision, night days K value (KD), psychology indicator (PSY), moving average convergence and divergence (MACD), and RSI have been applied in a rough set approach (Shen & Tzeng, 2014; Shen & Tzeng, 2015b). Besides, to determine the future share price predictability of Hong Kong, South Korea and Singapore, such features as 20 days simple moving average, 20 days exponential moving average, moving average crossover, MACD, Kaufman adaptive moving average, and the most optimised moving average have been used to analyse the increasing profitability (Phooi M'ng, 2018). For the Morgan Stanley Capital International Emerging Market, in terms of moving average, relative strength index, and moving average convergence divergence, the prediction of transaction costs, using technical analysis, provided evidence against the efficient market hypothesis for emerging market index (Metghalchi et al., 2019). Technical analysts always consider the influencing factors from various aspects. With the continuous increase in the number and spread of social web media, the stock market volatility is affected by information release, dissemination and public acceptance (Li et al., 2018).

*2.2. Machine Learning Techniques for Computational Finance*

To take full advantage of the strengths of advanced machine learning techniques to produce broader impacts, effective practical implementations of predictive systems must incorporate the use of innovative technologies. Stock prices prediction can be transferred to two types of problems: (1) decision making or classification problems for price trend prediction, such as fuzzy rule-based systems (ElAal et al., 2012), neural networks (ElAal et al., 2012, Lertyingyod & Benjamas, 2017), and random forests with imbalance learning (Zhang et al., 2018), and (2) time series prediction (TSP) problems for price value prediction. Various machine learning techniques have been applied for TSP problems (Jadhav et al., 2015, He & Qin, 2010). Especially, neural networks (e.g. recurrent neural networks and feed-forward neural networks (Chandra & Chand, 2016; Khare et al., 2017)) have been proven to be very promising for solving TSP problems in the literature.

Classic artificial neural networks (ANNs) have been widely used for predicting stock price (Khare et al. 2017, Göçken et al. 2019), digital content stocks (Chang, 2011), and the closing price of PETR4 stocks (Andrade De Oliveira et al., 2011). Various types of neural networks were developed for the stock price prediction, such as Bat-neural networks based on the quarterly data for eight-year DAX stocks (Hafezi et al., 2015), nonlinear autoregressive with exogenous inputs (NARX) neural networks based on the data from the NASDAQ stock AAPL (Wei & Chaudhary, 2016), Recurrent Neural Networks based on the data from NASDAQ stock exchange (Chandra & Chand, 2016), and Radial basis function (RBF) neural network on the average prices of previous 8 months (Liu, 2018),

Recently, deep learning techniques have been widely used for stock price prediction, for example, the deep direct reinforcement learning algorithm (Deng et al., 2017), Convolutional neural networks (CNN) (Cao & Wang, 2019) and deep Q-learning (Jeong & Kim, 2019). Support vector machine (SVM) was also widely used in this area. For example, Long et al. (2019) developed SVM based on different kernels to predict stock price trend based on based on financial news semantic and structural similarity.

Hybrid models always perform better than single machine learning models, and the neural networks and SVMs are often an important component in many hybrid models, for example, the combination of neural network and decision tree for the prediction of digital game content stocks price (Chang, 2011), the integration of gray algorithm and RBF neural network, trained by four different learning strategies (Lei, 2018), fuzzy time series analysis with neural networks for the forecast of the Taiwan Stock Exchange Capitalization Weighted Stock Index (Yolcu & Alpaslan, 2018), the integration of piecewise linear representation (PLR) and support vector machine (SVM) (Luo & Chen, 2013), and the combination of CNN and SVM (Cao & Wang, 2019).

Evolutionary optimization with wrapper techniques are often used to optimize machine learning models or select features for stock price (trend) prediction. For example, Harmony Search and Genetic Algorithm was used to find the best structure of neural network (Göçken et al., 2016), neural network with greedy algorithm based on the feature reduction with wrapper techniques for the prediction of stock price trend (Lertyingyod and Benjamas, 2017), simulated annealing algorithm was used to optimise feature space and model parameters (Torun & Tohumolu, 2011), and a Markov decision process was incorporated on genetic algorithms to develop stock trading strategies (Chang & Lee, 2017).

Many other techniques were used to improve the performance of computational finance. For example, principal component analysis (PCA), as data pre-processing, was used to reduce the dimensions of feature space to improve prediction accuracy (Lin et al., 2009), and it was applied to clean up the original data set and produce a new data structure (Chen & Hao, 2018), an analytic hierarchy process was used for feature ranking and selection for a weighted kernel LS-SVM (Marković et al., 2017), and the information gain oriented feature selection was run before the Genetic Algorithm with SVM in wrapper for credit scoring (Jadhav et al., 2018).

## 3. Methodology

*3.1. Problem Modelling*

This research focuses on the classic TSP problem for stock price prediction. It is to explore the advanced methods of share price prediction in the aerospace industry by using some historical finance data. A set of features, $x1, ..., x_j$, can be extracted from raw date to feed into a proposed model to predict the future close price, y.

Due to the uncertainty and dynamics of stock markets, the TSP problem is a strong non-linear regression problem. The value of the time series at time $t + \tau$ is predicted or forecast based upon the properties of the historical time series $t, t-1, ....., t-n+1$ (where $n$ is the number of time steps within the duration of the time series). The non-linear regression can be expressed as shown in Equation **(1)**.

$$y_{t+\tau} = f(x_t, x_{t-1}, ......, x_{t-n+1}) + \varepsilon_{t+\tau}, \tau \geq 1 \quad \text{(1)}$$

To solve this problem and accurately predict the future share price, two challenges need to be overcome. The first one is to choose proper features, and the other is to develop an appropriate model that represents the function *f(\*)*, mapping the non-linear relationship between the historical data and the future stock price to be predicted, and it should be robust for the uncertainty and complexity of data. There are hundreds of finance features for share price prediction and each feature has different influence levels in various industries, even has different influence levels for different companies in the same industry sector. The parameter $\tau$ could be different for different industries, due to their different data properties.

*3.2. Feature Selection and Data Processing*

After decades of stock research, more than 100 indicators and ratios have been developed for fundamental and technical analysis, respectively. In this research, data is collected from the Bloomberg databases, which is the most popular stock database. 30 features are selected, as they are the most popularly used in existing research. For the fundamental analysis, 15 features in 6 categories are selected, whereas for the technical analysis, 15 features in 10 groups are selected. The description and calculation formula of the 30 features are listed in Tables A1 and A2 in Appendix. The collected raw data will go through two stages of data processing.

For the fundamental features, data is collected every quarter after the companies publishing their financial reports. For the technical features, data is collected every trading day, synchronised with the close price. A sample is the data on every trading day. To compare the impact of two sets of features on technical analysis, the quarterly fundamental data is transformed to daily data by using the univariate linear processing, which is decided by the beginning and the ending values of the feature in the quarter, as shown in Fig. 1, where, assume all the change of a fundamental feature in a quarter is linear.

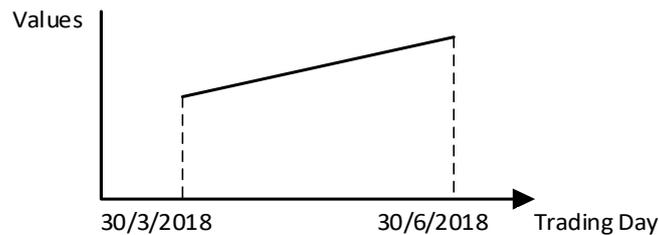

**Fig. 1.** Fundamental Features Processing Rule

When processing the technical data, to keep the integrity of the sample data, the samples that miss some feature values are deleted. At the same time, the corresponding trading day samples transferred from the fundamental data are also deleted to maintain the consistency of the fundamental data and technical data.

*3.3. Dimension Reduction with PCA*

To overcome the curse of dimensionality, the technique of classic Principal Components Analysis (PCA) is used. The idea of PCA is to project *n*-dimensional data onto *k*-dimensional (*n>k*) hyperplane, thus minimising the projection distance from each sample point to the hyperplane and maximise the variance. The implementation of PCA mainly includes five steps (Groth et al., 2013):

(i) Standardise the data samples by mean normalisation;
(ii) Calculate the covariance matrix of the data;
(iii) Find the eigenvalues and the eigenvectors of the covariance matrix above;
(iv) The obtained eigenvectors are combined according to the size of the eigenvalues to form a mapping matrix, and the largest number of the top k rows or the top k columns of the mapping matrix are extracted as the final mapping matrix;
(v) Mapping the original data with the mapping matrix of step (iv) to achieve the purpose of data dimensionality reduction.

## 3.4. Recurrent Neural Network (RNN)

### 3.4.1. The structure of an RNN

In the financial prediction problem, the current share price is highly related to the historical values of indicators and its historical values. RNN is a type of neural network that includes internal memory function through the feedback loops. It can store information while processing new inputs. This kind of memory makes it ideal for the tasks that need to consider history inputs, e.g. the TSP problem for share price. To predict the share price several days later, it may need to consider the features and share price a few days ago. Therefore, a target delayed RNN is proposed. The structure of target delayed RNN is shown in Fig. 2, where,

$x^{(t)}$ is a $j$-dimensional vector input ($j$ features) in the time-step $t$;

$h^{(t)}$ is the hidden state vector with $n$ hidden layers in the time-step $t$;

$y^{(t+\tau)}$ is the output on the $\tau^{th}$ day after current time $t$;

$\tau$ is the trading days of delay $1 < \tau \leq t$

$U_t$ is the direct weight from the input layer at time $t$ to the hidden layer at time $t$;

$V_t$ is the weight from the hidden layer at time $t$ to the output layer at time $t+\tau$;

$W_{t-1}$ is the weight from the output at time $t$-1+$\tau$ to the hidden layer at time $t$, which means the historical memory;

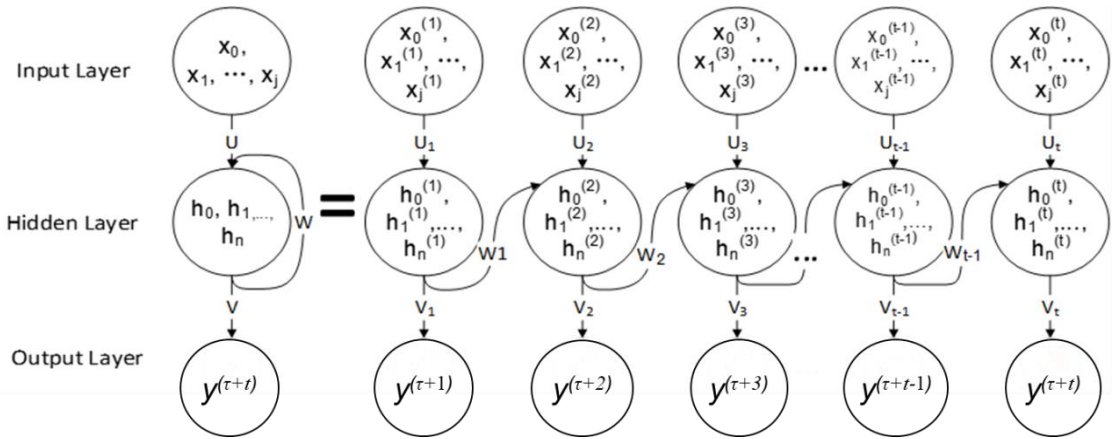

**Fig. 2.** The Structure of Target Delayed RNN

Therefore, the RNN output can be calculated in Eqs. (2) and (3)

$$h^{(t)} = f_h(U_t x^{(t)} + w_{t-1} h^{(t-1)}); \quad (2)$$

$$y^{(t+\tau)} = f_y(V_t h^{(t)} + b_y). \quad (3)$$

*3.4.2. The Training of the RNN*

There are many different algorithms for training an RNN. To obtain good performance of RNNs, three training algorithms are investigated, such as Levenberg-Marquardt Algorithm (Ranganathan, 2004), Bayesian Regularisation BP Algorithm (Mahajan et al., 2015; Burden & Winkler, 2008) and Scaled Conjugate Gradient Algorithm (Zhou & Zhu, 2014; (Møller, 1993).

**Levenberg-Marquardt** (LM) is the most widely used optimization algorithm. LM, as the damped least-squares method, is an estimation method for the regression parameter least squares estimation in the nonlinear regression. This method combines the Gauss-Newton algorithm and the Gradient Descent algorithm and converts the nonlinear least squares problem into a series of linear least squares problems, then solves the problem by iteration. Hence, it outperforms simple gradient descent and other conjugate gradient methods in a wide variety of problems. The details of LM is provided in (Ranganathan, 2004).

**Bayesian Regularisation** nets have been proved to be more robust than standard BP nets (Burden & Winkler, 2008). The idea of the BR Algorithm is to retain all the features but to avoid the excessive influence of a specific feature by reducing the underlying population parameter $\theta$, which obeys the Gaussian distribution, to improve the problem of overfitting. By using the Maximum A Posteriori (MAP) method to estimate the parameter $\theta$, and then obtain the cost function. Denote the prior distribution before the training set being trained and calculate the posterior distribution through the Bayesian formula. It is considered that the value of $\theta$ should maximise the MAP so that the cost function will be minimum. The details of the BR-BP algorithm was provided in (Mahajan et al., 2015).

The performance of the **Scaled Conjugate Gradient** (SCG) Algorithm is benchmarked against that of the standard BP algorithm (Møller, 1993). The basic idea of the SCG is to combine the conjugate property with the steepest descent method. It constructs conjugate directions of $\mathbf{n}$ two-conjugates by using the gradients at the known points. The minimum value in one direction is obtained by searching and optimising along the direction. It does not affect the minimum value in the obtained direction when searching the minimum value in other directions. After finding the minimum values for all $\mathbf{n}$ directions, the minimum value of the $\mathbf{n}$-dimensional problem is obtained. The details of SCG is provided in (Zhou & Zhu, 2014).

*3.5. Evaluation Method of Prediction*

The mean square error (MSE) on the testing data is used to evaluate the prediction accuracy of different models. The smaller the value of MSE, the better the accuracy of the prediction model.

$$MSE = \frac{1}{N}\sum_{i=1}^{N}(y(x^{(i)}) - y_{desired}^{(i)})^2 \qquad (4)$$

## 4. Experiments and Evaluation

*4.1. Data Processing*

The aim of this research is to analyse the share price of the aerospace industry. There are two kinds of companies in in the aerospace sector: the aerospace manufacturers and the aerospace operators. Two world's famous companies, MFG, one of the largest companies in aerospace manufacturing sector, and OPR, a famous airline holding company in the world, are analysed in this study.

The financial data of these two companies for the period of 5 years from 1st July 2013 to 29th June 2018, containing 1260 trading days, are collected from the Bloomberg database. As those samples missing some values in BA are removed, there totally are 1254 MFG samples for the experiments. All 1260 OPR samples are used for the experiments. In the experiments, 70% of the samples are used to train the model, 15% of samples are used for validation and 15% of samples are used for testing.

*4.2. Experiment Conduction*

The experiments are carried out on a laptop with Intel Core [i5-7200U@2.50GHz](i5-7200U@2.50GHz), RAM4.00GB and Windows 10. 216 experiments of RNN are conducted on the platform of MatLab with different sets of the features, training algorithms, neurons, delay days and target companies. A ten folder crossing validation is used for each experiment, taking 10 times of random training, validation and testing, and calculating the average performance (arithmetic mean) of the 10 times of tests. Fig. 3 depicts the experimental parameter setting.

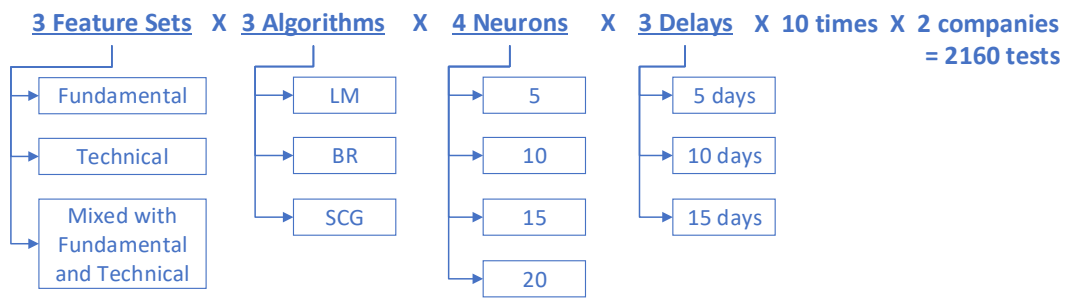

**Fig. 3.** The Parameters of Conducted Experiments

Before running RNN for prediction, all the datasets are processed with PCA to reduce the data dimensions. Six studies are implemented in Python on the PyCharm platform with 15 fundamental features, 15 technical features and 30 mixed features of data for MFG and OPR. Two and three components are kept in each study for easy visualisation of PCA results.

The two termination criteria and the learning rate of the developed RNN are set as below:
1) The training ends when it reaches convergence;
2) If it does not converge within 1000 epochs, automatically ends at the 1000th epoch;
3) The learning rate is 0.02; the number of hidden layer is 1.

*4.3. Results on MFG*

*4.3.1. Data Dimensionality Reduction with PCA*

Fig. 4 (1) – (3) illustrate the results of PCA on the MFG data. Fig. 4 (1) plots several lines connected to form the 2D and 3D graphs for fundamental data, while Fig. 4 (2) and (3) are scatter plots for technical features and mixed features, respectively. This is because that fundamental data is transferred from quarterly to daily through univariate linear function, while technical data is the actual daily financial data gathered from the database. Fig. 4 (3) is similar to Fig. 4 (2) in both 2D and 3D graphs. This indicates that the fundamental data of MFG has a small influence on the results of PCA, when it is combined with technical data.

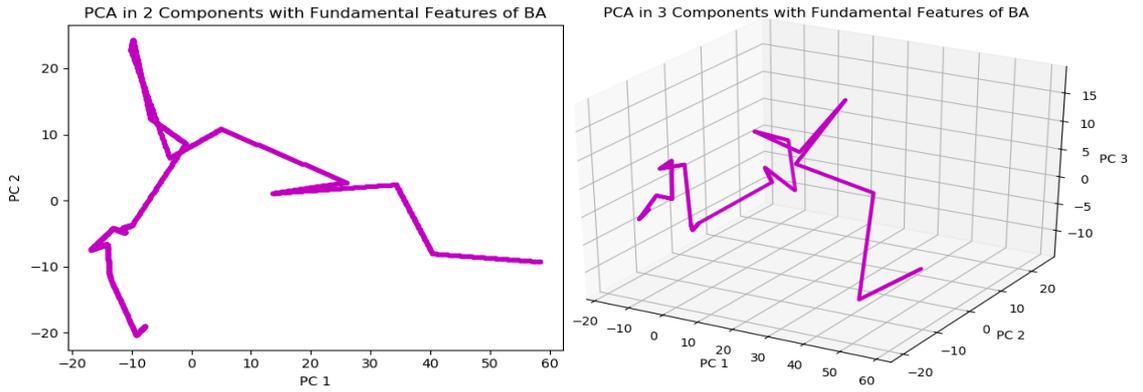

(1) PCA components on Fundamental Features of MFG

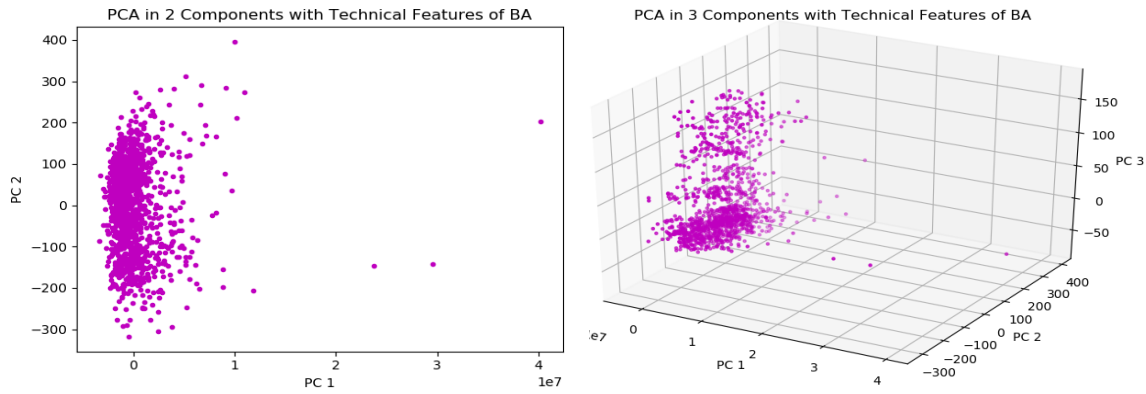

(2) PCA components on Technical Features of MFG

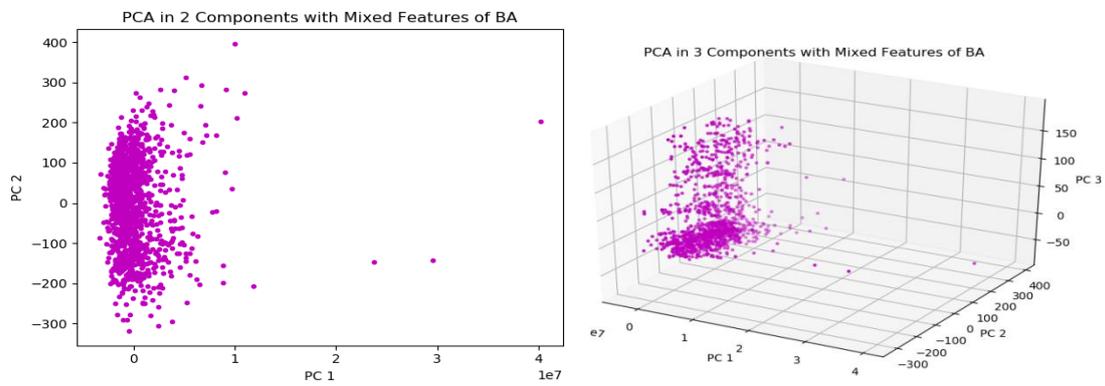

(3) PCA components on Mixed Features of MFG

**Fig. 4.** 2D and 3D Visualisation of PCA components

In PCA, the loss of detailed information is inevitable. The variance explained ratio is used to indicate the cumulative contribution rate of the first few principal components extracted by the PCA. Table 1 shows the variance explained the ratio for MFG.

**Table 1**

The Explained Variance of Principle Components on the MFG data

| PC | The Explained Variance of Principle Components | | | | | |
|---|---|---|---|---|---|---|
| | Fundamental | | Technical | | Mixed | |
| | Individual | Cumulative % | Individual | Cumulative % | Individual | Cumulative % |
| 1 | 0.6606753 | 66.07% | 9.99999997e-01 | 99.99% | 9.99999997e-01 | 99.99% |
| 2 | 0.2303135 | 89.10% | 2.00097504e-09 | 99.99% | 2.00243295e-09 | 99.99% |
| 3 | 0.0684519 | 95.94% | 4.97881623e-10 | 99.99% | 5.48986281e-10 | 99.99% |

*4.3.2. Convergence of RNN training*

Fig. 5 illustrates the training process of RNN with BR. It can be seen that the MSE falls to the lowest position sharply at epoch 10, and then the error almost does not change. The training process continues until epoch 683, with the best performance 4.2123 at epoch 675. Through observing all the training processes of RNN with BR, it is shown that BR always converges very quickly, but the adjustment process is slow. Namely, the MAP in the Bayesian algorithm finds a set of approximate optimal parameters very fast, and then the fine adjustment of weights is slowly.

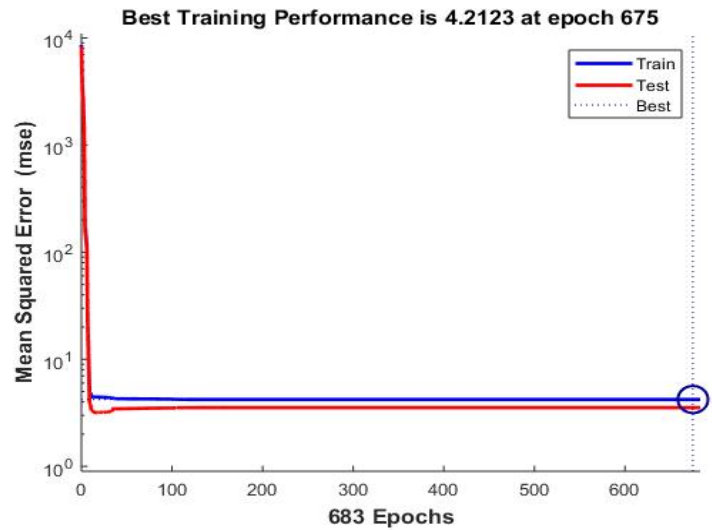

**Fig. 5.** The MSE evolution of the best RNN on the MFG training data

*4.3.3. The prediction results of RNN on the MFG data*

The most accurate prediction of the share price of MFG is produced by the RNN on the experimental parameters: "BR Algorithm, Mixed Features, 15 Neurons and 5 Delays" with the average MSE of 3.482006. Fig. 6 shows that the errors are increased from time-step 1000 to 1200. It means that although the RNN obtained by BR has the best performance, the accuracy is reduced from June 16 2017 to April 4 2018. Namely, the dynamics of the stock market prices has changed very much in the past year.

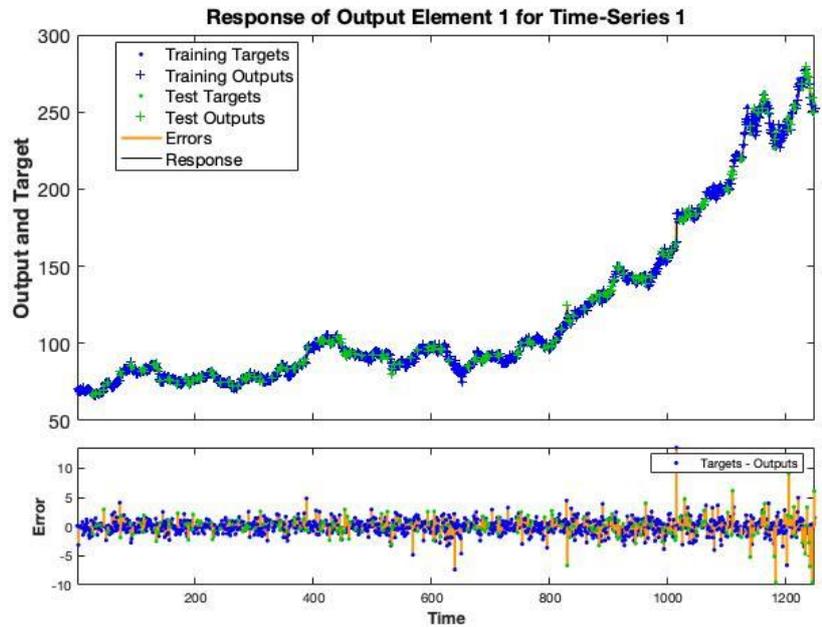

**Fig. 6.** The outputs and errors of the best RNN on the MFG test data

*4.4. Results on OPR*

*4.4.1. Data Dimensionality Reduction with PCA*

Fig. 7 (1) shows a unique trend for the fundamental data of OPR, which is different to Fig. 4 (1) for the fundamental data of MFG, and Fig. 7 (2) looks more compact for the technical data of OPR than Fig. 4 (2) for the technical data of MFG. Furthermore, the most significant difference is the result of PCA on the mixed data of OPR. As can be seen in Fig. 7 (3), the data in the 3D graph is divided into two clusters for the mixed data of OPR, deferring to the 3D graph in Fig. 4 (3) for the mixed data of MFG. This means both fundamental data and technical data affected the results of PCA for OPR. Namely, fundamental data could play different roles for the stock price prediction in aerospace manufacturer and operator.

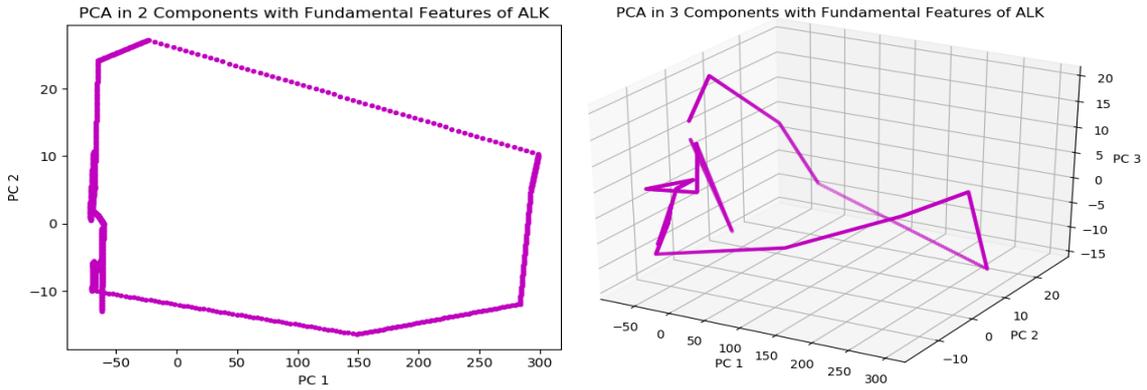
(1) PCA with Fundamental Features of OPR

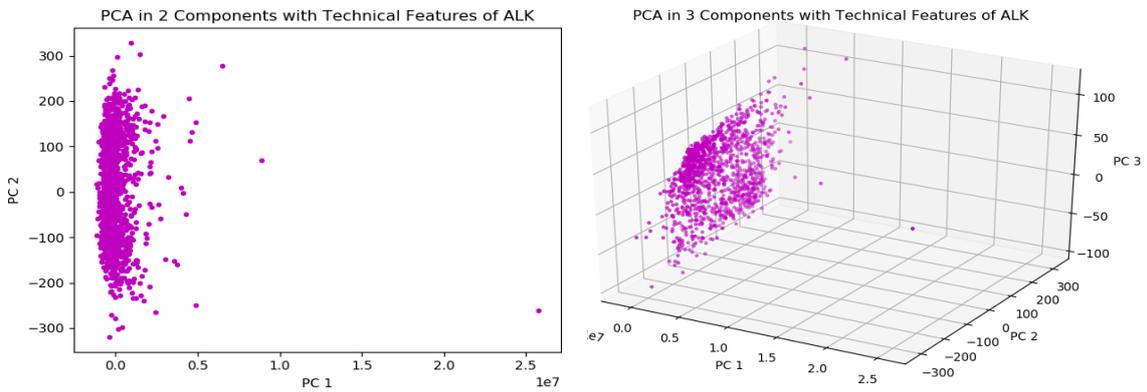
(2) PCA with Technical Features of OPR

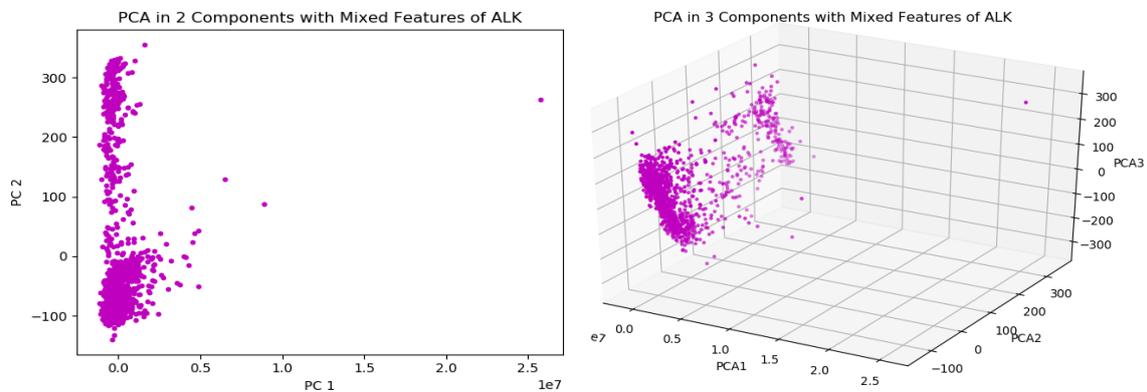
(3) PCA with Mixed Features of OPR

**Fig. 7.** 2D and 3D visualisation of PCA components

Table 2 shows that 99.82% of details can be represented with 3 components for OPR. The percentage for OPR is larger than the percentage of 95.94% for MFG. It should be noticed that technical data and mixed data can be explained close to 100% in the first component. This means the data with 3 components after PCA can correctly explain all the raw data with 15 or 30 features. In addition, the results of PCA on the mixed data are similar with that on the technical data for both MFG and OPR. This means that the technical data has more influence on the results of PCA on the mixed data.

**Table 2**

The Explained Variance of Principle Components on the OPR data

| PC | The Explained Variance of Principle Components | | | | | |
|---|---|---|---|---|---|---|
| | Fundamental | | Technical | | Mixed | |
| | Individual | Cumulative % | Individual | Cumulative % | Individual | Cumulative % |
| 1 | 0.9880090 | 98.80% | 9.99999988e-01 | 99.99% | 9.99999974e-01 | 99.99% |
| 2 | 0.0071197 | 99.51% | 1.08145916e-08 | 99.99% | 1.35294546e-08 | 99.99% |
| 3 | 0.0030929 | 99.82% | 1.19720806e-09 | 99.99% | 1.06595881e-08 | 99.99% |

*4.4.2. The convergence of RNN Training*

Fig. 8 is the convergence curve of the best RNN on OPR. The convergence speed of RNN, trained by the SCG algorithm with the ending epoch at 33 for OPR, is faster than that for MFG, and the downward trend is relatively flat. This result shows that the SCG algorithm, combined gradient descent and conjugation, does not need a one-dimensional search that costs time.

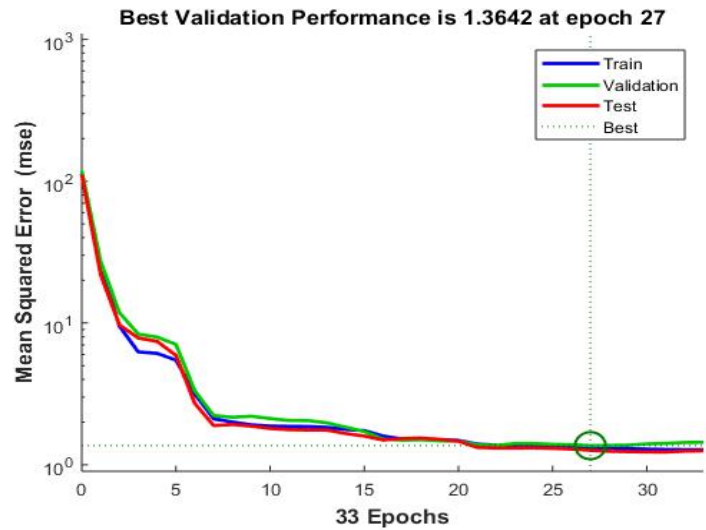

**Fig. 8.** The MSE evolution of the best RNN on the OPR training data

*4.4.3. Results of RNN on the OPR data*

For OPR, the best RNN is produced with the experimental parameters: "SCG Algorithm, Mixed Features, 5 Neurons and 15 Delays" with the average MSE of 1.319737. Fig. 9 shows the model outputs and errors for OPR. The error for OPR is evenly distributed and smaller than that for MFG, as in the SCG algorithm, every direction has a conjugated direction and all the directions could be searched.

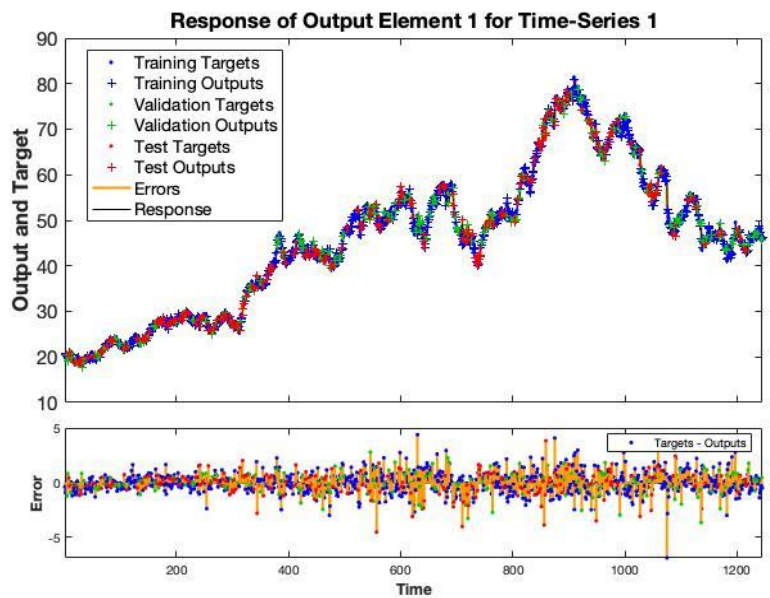

**Fig. 9.** The Outputs and Errors of the Best RNN on the OPR Test Data

*4.5. Evaluation*

The best results of the PCA and RNN for the share price prediction of MFG and OPR have been shown in the last section. The experimental outcomes will be further assessed with more dimensions as follows.

*4.5.1. Principal Components Analysis*

The benefits of PCA lie in the running time reduction, data visualisation and compression to free up the computer memory space. The most important is the benefit of reducing the computing complexity, when a large number of features is dealt with for the specific problem. In this research, although there are only 15 (fundamental, technical) or 30 (mixed) features, the average running time of each test can be from 1 second to maximum 92 seconds. Hence, PCA is necessary for this model.

Firstly, the running time of RNN depends on its structure. When the number of hidden neurons and delays of RNN is increased, its running time is increased. The total running time of data processing, results extraction and analysis of the 2160 tests are around 6 days within only 5-years financial samples. When expanding the feature sets or adding the financial samples, the running time could be increased sharply.

Besides the improved efficiency, the PCA can keep more than 95.94% details, but the number of data dimensions is reduced from 15 or 30 to 3. For the technical and mixed features of both MFG and OPR, more than 99.99% variance is explained with one-component. This means that data redundancy is reduced, and the significant information of data is kept.

Additional 20 verification tests for the two test sets are done to examine the effects of PCA further. The parameters are set the same as the best-performed tests of MFG and OPR:

- For MFG: BR Algorithm, Mixed Features, 15 Neurons and 5 Delays
- For OPR: SCG Algorithm, Mixed Features, 5 Neurons and 15 Delays

But, the input data of RNN is the initial data without PCA processing. Tables 3 and 4 show the assessment results for MFG and OPR, respectively. From both tables, the average MSE with PCA is lower than average MSE without PCA. This means that the prediction accuracy with PCA is better than without PCA. Moreover, the average running time with PCA is lower than or similar to that without PCA. The results have proved that PCA can improve both efficiency and accuracy of the prediction model.

**Table 3**

PCA Assessment Result of MFG

|             | Average Total Epochs | Average Best Performance Epochs | Average MSE | Average Running Time (s) |
|-------------|----------------------|----------------------------------|-------------|--------------------------|
| With PCA    | 614.7                | 533.9                            | 3.482006    | 5.4                      |
| Without PCA | 764.4                | 649                              | 4.013047    | 6.4                      |

**Table 4**

PCA Assessment Result of OPR

|             | Average Total Epochs | Average Best Performance Epochs | Average MSE | Average Running Time (s) |
|-------------|----------------------|----------------------------------|-------------|--------------------------|
| With PCA    | 49.4                 | 43.4                             | 1.319737    | 1                        |
| Without PCA | 76.8                 | 70.8                             | 7.517148    | 1                        |

*4.5.2. Optimization Algorithms of Weights*

BR and LM account for the majority of the top 50 test sets for MFG, while SCG and LM perform better than BR for OPR. For the MFG data, LM obtains the lowest average MSE and running time among the three algorithms. Although the best performance is obtained by BR algorithm, the average MSE obtained by BR is higher than that obtained by LM, and the average number (773.1) of convergent epochs for BR is the largest among the three algorithms.

**Table 5**

Testing Results of Optimization Algorithms of MFG

| Algorithms | Average Total Epochs | Average Best Performance Epochs | Average MSE | Average Running Time (s) |
|---|---|---|---|---|
| LM | 21.4 | 15.4 | 5.445273 | 1.0 |
| BR | 773.1 | 714.3 | 10.369097 | 19.5 |
| SCG | 60.6 | 54.6 | 9.581278 | 1.0 |

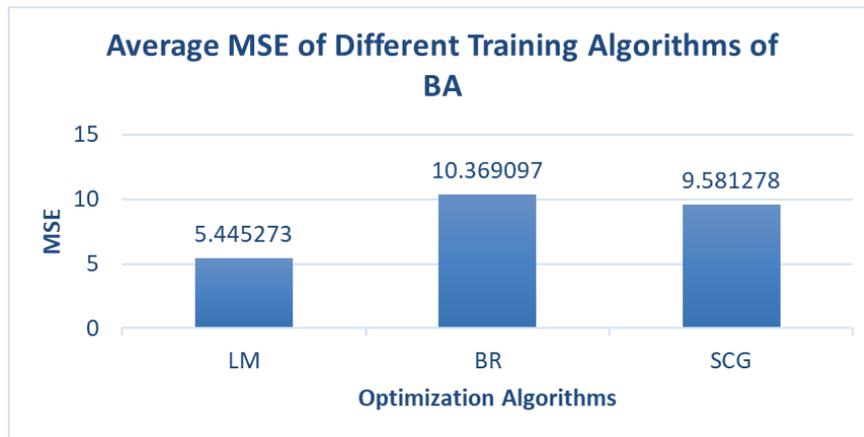

**Fig. 10.** Average MSE of Different Training Algorithms of MFG

For OPR, the best algorithm is SCG with the average MSE of 4.057354 and average running time of 1 second.

**Table 6**

Testing Results of Optimization Algorithms of OPR

| Algorithms | Average Total Epochs | Average Best Performance Epochs | Average MSE | Average Running Time (s) |
|---|---|---|---|---|
| LM | 15.7 | 9.7 | 5.374760 | 1.0 |
| BR | 859.5 | 777.2 | 6.981070 | 19.7 |
| SCG | 65.9 | 59.9 | 4.057354 | 1.0 |

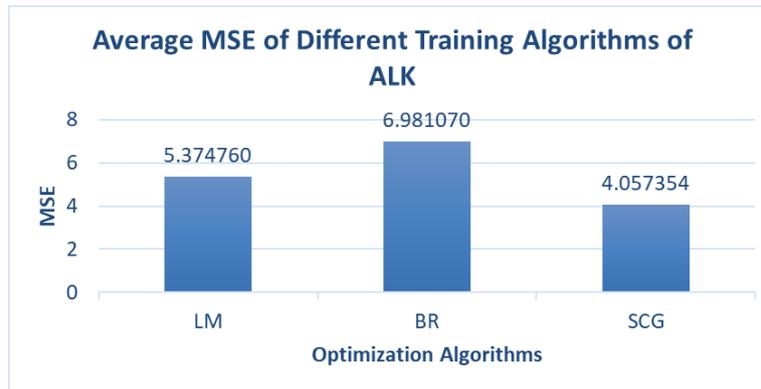

**Fig. 11.** Average MSE of Different Training Algorithms on the OPR Data

When analysing the different algorithms separately, the BR algorithm is the only one that does not converge within 1000 epochs when testing. Especially in the tests for OPR, 33 of 36 tests with BR are ended at the 1000 epoch, while this number is 12 of 36 in the test for MFG. All these tests are carried out, using the RNN models with over 15 hidden neurons and over 10 delays. This indicates that the threshold value of convergence may not be suitable for the BR algorithm with many hidden neurons and delays.

To sum up, from the average value of MSE and running time, LM is the best algorithm for MFG while SCG is the best one for OPR. Even BR performs very well in the single test for MFG, it costs a lot of running time and has a high average MSE when the convergence threshold of BR is set to the same threshold as in LM and SCG. Hence, the convergence threshold should be refined.

*4.5.3. Analysis Features*

The experiments on the technical analysis features obtain the lowest average MSE for MFG, regardless of algorithms, and the experiments on the fundamental analysis features obtain the best for OPR, as shown in Fig. 12 and Fig. 13.

**Table 7**

Testing Results on the Three Groups of Features for MFG

| Features | Average Total Epochs | Average Best Performance Epochs | Average MSE | Average Running Time (s) |
|---|---|---|---|---|
| Fundamental | 287.4 | 262.2 | 8.401831 | 7.0 |
| Technical | 283.0 | 263.2 | 8.257568 | 7.0 |
| Mixed | 284.6 | 258.8 | 8.736249 | 7.5 |

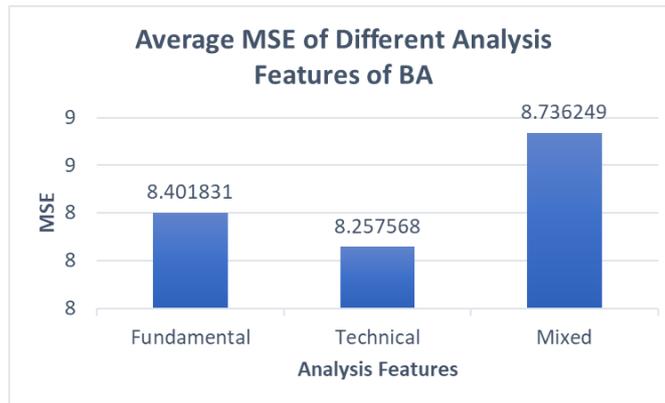

**Fig. 12.** Average MSE of Different Analysis Features of MFG

**Table 8**

Testing Results of Analysis Features of OPR

| Features | Average Total Epochs | Average Best Performance Epochs | Average MSE | Average Running Time (s) |
|---|---|---|---|---|
| Fundamental | 306.0 | 269.2 | 5.399674 | 7.2 |
| Technical | 314.9 | 280.4 | 5.588128 | 6.6 |
| Mixed | 320.2 | 297.1 | 5.425381 | 7.9 |

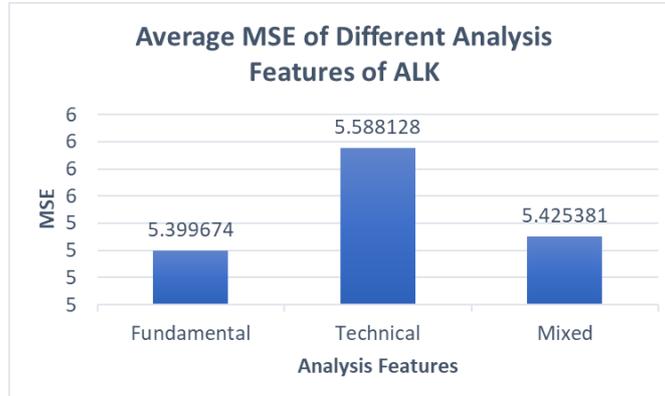

**Fig. 13.** Average MSE on The Three Sets of Features for OPR

For the MFG Company, as the dynamics of information is approximated as a linear variation, fundamental data misses details of daily information, so the accuracy of the prediction is not as good as the technical data, which represents daily information. Besides, even though fundamental data occupies a small percentage of the influence on mixed data, these filled fundamental features may conflict with the technical features, thus producing negative impact on the prediction of stock price. Hence, the prediction accuracy of the model on mixed features is reduced.

However, for the OPR Company, the stability of historical share price is poor, as shown in Fig. 9. That will cause a high fluctuation of the prediction accuracy if the technical data is used. The prediction accuracy can be improved and the high fluctuation could be mitigated by using mixtures of fundamental and technical features. This indicates the quarterly data has the positive impact on the prediction accuracy when the stock market is strongly dynamic. This might be because that the linear approximation of daily information from quarterly data could compromise with the dynamics of technical daily data.

Briefly, technical features can be selected when the share price is stable, while fundamental features are better when the share price has a high fluctuation.

*4.5.4. The Number of Neurons*

Many parameters can be adjusted for improving the prediction accuracy of a neural network. Appropriately adjusting the parameters plays a vital role in share price prediction. The method of trial and error is used to find the most suitable parameter values of RNN on different feature sets for different companies.

Fig. 14 shows the relationship between the number of neurons and the average MSE on the test data for MFG. The value of the average MSE increases for testing data as the number of neurons increases. The best-performed number of neurons is 5. This might be because that the more the neuron numbers, the stronger of the neural network's learnability, which might increases the risk of over-fitting with the training data.

**Table 9**

Testing Results for Different Numbers of Neurons in RNN for MFG

| Neurons | Average Total Epochs | Average Best Performance Epochs | Average MSE | Average Running Time (s) |
|---|---|---|---|---|
| 5 neurons | 241.9 | 219.6 | 6.839601 | 1.8 |
| 10 neurons | 268.5 | 246.6 | 8.192021 | 3.6 |
| 15 neurons | 312.6 | 285.5 | 9.193409 | 8.5 |
| 20 neurons | 317.0 | 294.0 | 9.635832 | 14.8 |

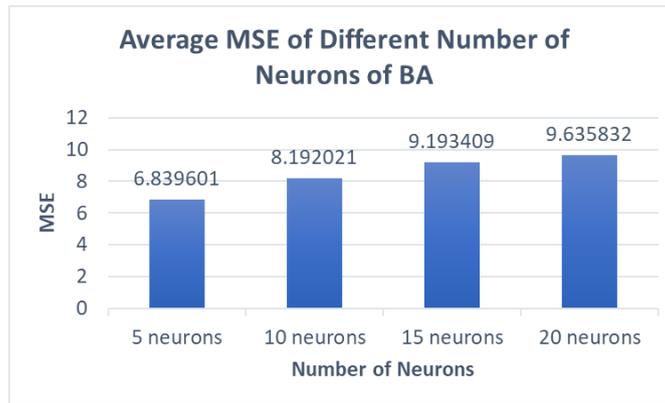

**Fig. 14.** Average MSE for Different Numbers of Neurons in RNN for MFG

As to the OPR, the MSE continues decreasing when the number hidden neurons is changed from 5 to 20. With the increasing number of neurons, the learnability of the RNN is improved, but the complexity of the network is increased. It can be seen that the MSE value has a tiny change when the number of hidden neurons is changed from 15 to 20. Therefore, the best-performed RNN is with 15 hidden neurons for OPR.

**Table 10**

Testing Results of Different Number of Neurons of OPR

| Neurons | Average Total Epochs | Average Best Performance Epochs | Average MSE | Average Running Time (s) |
|---|---|---|---|---|
| 5 neurons | 303.8 | 275.0 | 6.255731 | 2.3 |
| 10 neurons | 317.9 | 298.4 | 5.401003 | 4.5 |
| 15 neurons | 318.6 | 293.3 | 5.124511 | 8.2 |
| 20 neurons | 314.5 | 262.4 | 5.103000 | 14.0 |

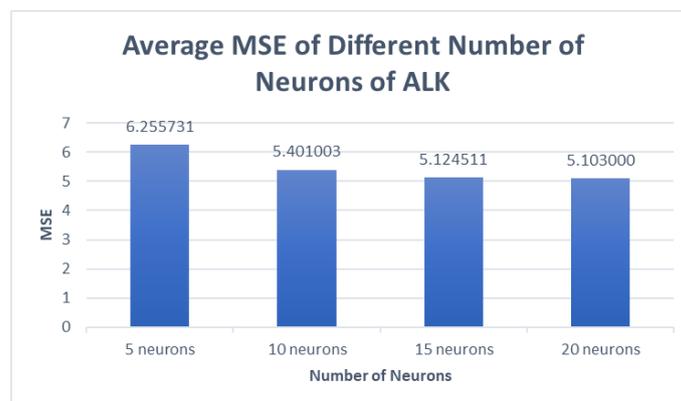

**Fig. 15.** Average MSE of Different Number of Neurons of OPR

Briefly, it is vital for improving prediction accuracy to set the number of hidden neurons appropriately. The experimental results indicate that there exists an optimal number of hidden neurons for RNN, and the number of hidden neurons depends on the nonlinearity of the training data. When the linearity of the training data is strong, the number of hidden neurons could be small, and when the nonlinearity of the training data is strong, the number of neurons could be large. When the number of hidden neurons is not enough, the trained model cannot well represent the mapping function between the data and the predicted value, the prediction accuracy will be not good enough. When the number of neuron is larger than the best number, the prediction accuracy on the test data set could be dropped, as the trained model might over-fit with the noise in the training data.

*4.5.5. The Number of Delays*

The delays is another important parameter of RNN to be adjusted for improving the performance of RNN. It also determines the time looking back of the history data for the prediction of share prices. As to the financial analysis, the most commonly used indicators are 5-days moving average, 10-days moving average and 15-days moving average. Correspondingly, the delays are set to 5, 10 and 15 delays. From Table 11 and Fig. 16, it can be seen that the MSE of RNN outputs for MFG is increasing as the number of delays of RNN is increasing.

**Table 11**

The Performances of RNN for Different Delays for MFG

| Delays | Average Total Epochs | Average Best Performance Epochs | Average MSE | Average Running Time (s) |
|---|---|---|---|---|
| 5 delays | 243.8 | 215.5 | 5.920882 | 2.3 |
| 10 delays | 301.0 | 281.8 | 9.397919 | 6.1 |
| 15 delays | 310.2 | 286.9 | 10.076847 | 13.1 |

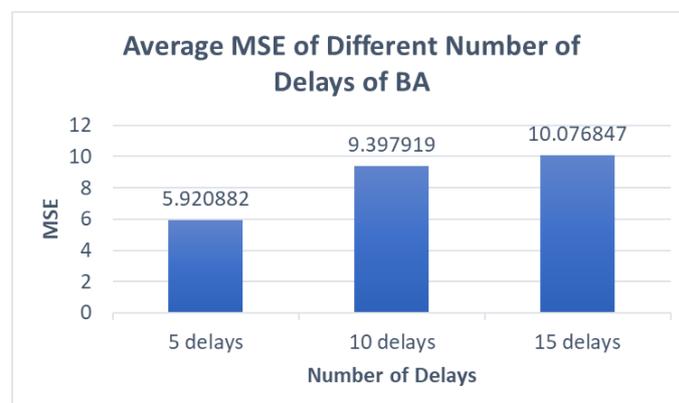

**Fig. 16. The** Average MSE of Prediction for Different Delays of RNN for MFG

However, the prediction accuracy of RNN for OPR has the opposite trend, compared to the case of MFG. The MSE is decreasing, as the number of delays is increasing for OPR (Fig. 17). Similar with the case of MFG, the running time of RNN is increasing, due to the increasing complexity of RNN (Table 12).

**Table 12**

The Performance of RNN for Different Delays for OPR

| Delays | Average Total Epochs | Average Best Performance Epochs | Average MSE | Average Running Time (s) |
|---|---|---|---|---|
| 5 delays | 294.3 | 245.2 | 5.935670 | 2.8 |
| 10 delays | 320.1 | 294.2 | 5.593137 | 6.2 |
| 15 delays | 326.6 | 307.3 | 4.884377 | 12.7 |

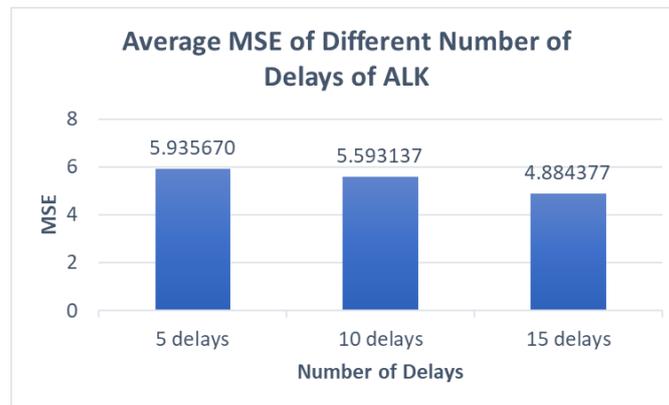

**Fig. 17. The** Average MSE of Prediction for Different Delays of RNN for OPR

Compared Fig. 16 with Fig. 17, for MFG, the short-term historical data could have positive impacts on its share price because the historical data is stable and do not need to consider long-term historical data by data processing of fundamental data. In contrast, for OPR, the long-term historical data could have positive impacts on its share price, as the long-term historical data could compensate for the high fluctuation of short-term historical data of OPR.

In brief, the relationship between MSE and the number of delays varied with the stability of the historical data of different companies. Hence, the delays should be adjusted to the optimal value for different companies.

**5. Conclusions**

The research developed a hybrid approach to predicting share price of aerospace industry companies by using the combination of PCA and RNNs, trained with LM, BR, and SCG algorithms, respectively. The proposed approach could help create automatic prediction of share prices for different industries, not only aerospace industry sectors, but also other industry sectors.

The share prices for two types of aerospace industries, manufacturing and operating companies, are investigated, and a comprehensive assessment is provided. Various factors, e.g. training algorithms, structure of RNN, and features, could affect the prediction performance. The experimental results show that PCA could improve the performance of prediction not only in efficiency, but also in prediction accuracy. For MFG, the best prediction accuracy was obtained by the RNN with 15 hidden neurons and 5 delays on the mixed features, optimised by the BR Algorithm, whereas, for OPR, the best prediction accuracy was obtained by the RNN with 5 hidden neurons and 15 Delays on the mixed features, optimised by the SCG algorithm. Regarding the robustness, LM obtained the best average MSE and shortest running time for MFG.

More importantly, the experimental results show that feature selection is related to the stability of share market. Technical features could be selected when the share price is stable, whereas fundamental features are better when the share market has a high fluctuation. The delays of RNN are different for different types of companies. It would be more accurate through using short-term historical data for MFG, whereas using long-term historical data for the OPR Airline.

The developed approach to predicting stock prices could provide decision-makers with a reference or evidence for their economic strategies and business activities, thus helping the financial industry to increase the return on investment. Although the research was done based on the data in aerospace industry at pre-COVID-19 time, the developed approach can be used for the share price analysis of aerospace industry at post COVID-19 time.

The combination of fundamental and technical analysis in this research fills the gap in the studies of stock prices that used to use only fundamental or technical analysis individually. However, simply combination of technical and fundamental data may not be good enough for the prediction of stock price for aerospace industries. Hence, feature selection with optimisation algorithms will be future work. Currently, COVID-19 is severely influencing aerospace industry, so, we will investigate the prediction of aerospace share price at post-COVID-19 time.

**Acknowledgements**

Zhou, F., & Zhu, X. (2014). Alphabet recognition based on Scaled Conjugate gradient BP algorithm. *Lecture Notes in Electrical Engineering*. https://doi.org/10.1007/978-3-642-40640-9_3
34

# Appendix

## Table. A1.

Fundamental Features

| Categories | Features | Descriptions | Formulas |
|---|---|---|---|
| Profitability Ratios | Return on Invested Capital | ROIC is the ratio between the funds invested and the related returns. It is used to assess the historical performance of a company or its business unit. | EBIT x (1 - tax rate) / (interest bearing liabilities + shareholders' equity - cash and its equivalent) |
| | Operating Margin | Operating Margin is the ratio of operating profit to operating income of the company | operating profit / operating income * 100% |
| | Return on Assets | ROA is an indicator used to measure how much net income is generated per unit of assets. It is used to assess the profitability of a company relative to its total asset value. | net income / total asset |
| Growth | Revenue Growth | Revenue growth is an indicator to measure the growth of revenue. Here the calculation period is one year. | revenue - revenue of last year |
| | Total Assets Growth | Total assets growth is an indicator to measure the growth of total assets. Here the calculation period is one year. | total assets - total assets of last year |
| | Total Debt Growth | Total debt growth is an indicator to measure the growth of total debt. Here the calculation period is one year. | total debt - total debt of last year |
| Leverage Ratios | Debt/Assets | The debt/assets ratio is a measure of the ability of an enterprise to use creditors to provide funds for its business activities and reflects the security level of creditors' loans. | total debt/total assets |
| | Equity Ratio | It is the ratio of shareholders' equity to total assets. This ratio reflects how much of the company's assets are invested by the owner and also shows the use of the company's financial leverage. | Equity/Total Assets |
| Efficiency Ratios | Accounts Receivable Turnover | The accounts receivable turnover rate is the average number of times that the company's accounts receivable are converted into cash within a specified period. | current sales net income / average balance of accounts receivable |
| | Quick Ratio | The quick ratio is one of the indicators for measuring the liquidity of a company's assets, reflecting the ability of the company's cash or immediately realisable assets to repay their current liabilities. | (current assets - inventory) / current liabilities |
| | Current Ratio | The current ratio is the ratio of current assets to current liabilities. It is used to measure the ability of a company's current assets to become cash for repayment of liabilities before the short-term debt expires. | current assets / current liabilities |
| Earnings | Earning Per Share | EPS is the net profit or net loss per share of the company. It is often used to reflect the company's operating results, measure the profitability of common stocks and investment risks. | (gross profit for the period - Dividend of preferred shares) / Total equity at the end of the period |
| | Price Earnings Ratio | The P/E ratio reflects the fact that when the dividend per share is unchanged, and the dividend pay-out ratio is 100% and the dividends received are not reinvested, how many years the investment can be fully recovered through dividends. | current market price per share/earnings per share of the period |

**Table. A2.**

Technical Features

| Categories | Features | Descriptions | Formulas |
|---|---|---|---|
| Volume | Volume | The total amount of securities or contract transactions during a specific period. | The sum of all the stock trade |
| Bollinger Bands | BB Width | Bollinger Bands uses statistical principles to determine the standard deviation of stock prices and their confidence intervals to determine the range of stock price fluctuations and future trends. It uses the waveband to display the safe, high and low price of the stock price. The upper and lower limits are not fixed and change with the rolling of the stock price. %b is mainly the position of the current price in the Bollinger Band, which is a crucial indicator when making trading decisions. | 100 * ((upper band price - lower band price) / 20 days moving an average of price) |
| | %B | | (close price - lower band price) / (upper band price - lower band price) |
| Commodity Channel Index | CCI | The CCI indicator specifically measures whether stock price, foreign exchange or precious metal trading have exceeded the normal distribution range. | ((high price + low price + close price) / 3 - average of 13 days close price) / ( average of 13 days close price minus close price * 0.015) |
| Rate of Change | ROC | ROC is the price of the day compared to the stock price of a specific day before a certain number of days. The speed of change reflects the degree of change in the stock market. | (current price - the price of 1 day ago) / price of 1 day ago |
| Relative Strength Index | RSI | RSI calculates the strength of market movements by calculating the magnitude of the stock price fluctuations and predicts the trend's persistence or turn. | 100 - [ 100 / (1 + (14 days smoothed moving average of up closes / 14 days smoothed moving average of down closes))] |
| Directional Movement Index | +DMI | The DMI indicator is a technical indicator that provides a basis for judging the trend by analysing the changes in the equilibrium points of the buyers and sellers during the ups and downs of the stock price. It has three indicators: +DMI, -DMI and ADX | 100 * (14 days smoothed moving average of + DM) / 14 days smoothed moving average of the true range |
| | -DMI | | 100 * (14 days smoothed moving average of - DM) / 14 days smoothed moving average of the true range |
| | ADX | | 100 * (14 days smoothed moving average of DX) where DX = 100 * |
| | | | abs((+DMI)-(-DMI))/((+DMI)+(-DMI)) |
| Simple Moving Average | SMA | SMA refers to the simple averaging of the closing price of a specific period. | SMA (5) = [price (1) + price (2) + ... + price (5)] / 5 |
| Moving Average Convergence and Divergence | MACD | The MACD is a technical indicator for judging the timing of buying and selling by using the short-term (usually 12-days) exponential moving an average of the closing price and the long-term (typically 26-day) exponential moving average. | 12 days exponential moving average - 26 days exponential moving average |
| | MACD Signal Line | | 9 days exponential moving average of MACD Line |
| KDJ Index | %K | KDJ calculates the immature random value RSV of the last calculation period by the highest price, the lowest price and the closing price of the last calculation period and the proportional relationship between the three. Then calculate the K value and the D value according to the method of smoothing the moving average and draw a graph to judge the stock trend. | 100 * (close price - range minimum) / (range maximum - range minimum) |
| | %D | | 3 days moving an average of %K |
| Williams %R | WR | WR is a technical indicator that uses the oscillation point to reflect the phenomenon of overbought and oversold in the market. It predicts the highs and lows in the cycle and then proposes active signals to analyse the market's short-term market trends and determines the market's strength and weakness. | 100 * (14 days highest price - close price) / (14 days highest price - 14 days lowest price) |